\documentstyle[pra,twocolumn,aps]{revtex}
 
\draft 

\title{Conditional generation of error syndromes in fault-tolerant
error correction}
\author{M.B. Plenio, V. Vedral, P.L. Knight
    \\ Blackett Laboratory,
    Imperial College, \\Prince Consort Road, London SW7 2BZ, U.K.}

\date{\today}

\begin{document}
\maketitle

\begin{abstract}
In this paper we show how the fault--tolerant error correction scheme 
recently proposed by DiVincenzo and Shor may be improved. Our scheme, 
unlike the earlier one, can also deal with a single error that might 
occur {\em during} the gate operations that are required for the 
implementation of the error correction and not only in--between the gates 
and hence presents an improvement towards enabling error correction
and with it the practical possibility of some more involved quantum 
computations, such as e.g. factorization of large numbers.
\end{abstract}

\pacs{03.65.-w, 89.70.+c, 89.80.+h}

Soon after the idea of quantum computation became an active part of 
current research through the seminal work of Shor on factorization 
\cite{Shor1}, decoherence and especially spontaneous emission was 
recognized as a major problem that 
cannot be ignored, at least when one is interested in practical 
applications including especially the factorization of large numbers 
\cite{Plenio1,Plenio2}. It has become clear that efficient error 
correction methods have to be found to overcome decoherence if we are
ever to 
realize the possibility of building a quantum computer. In fact, 
inspired by the theory of classical error correction \cite{Pless1}, 
Shor, Calderbank and Shor and independently Steane 
\cite{Shor2,Steane1,Steane2,Calderbank1} proposed the first quantum error 
correction codes able to correct errors that occur during the 
{\em storage} of qubits. More error correction codes have been discovered 
and theoretical work undertaken has elucidated the structure 
of quantum codes 
\cite{Ekert1,Laflamme1,Bennett1,Chuang1,Plenio3,Vaidman1,Knill1,Barenco1,
Calderbank2,Shor3,Steane3,DiVincenzo1,Gottesmann2,Gottesmann1,Cleve1,Calderbank3}
even further.

However, these investigations have dealt with the problem of processing 
the stored information and not with the correction of errors that might 
occur during gate operation. A significant step forward in this direction 
was made by Shor and DiVincenzo \cite{Shor3,DiVincenzo1}. In \cite{Shor3} 
the idea of {\em fault--tolerant} implementation of quantum gates was 
developed where one error would not lead to many errors,  and hence can 
be corrected by subsequent error correction. An example of a fault--tolerant 
network for error correction proposed in \cite{DiVincenzo1} is shown in 
fig.1. It has the
important property of being able to perform error correction even if an
error occurs between the execution of two of its quantum gates. To be more 
precise, if the incoming state is error free, and one error occurs during 
the error correction, then the outgoing state will have {\em at most} one 
error. This is a substantial improvement compared to previous error 
correcting networks, where one error during error 
correction usually results in more than one error in the outgoing state. 
However, the network given in \cite{DiVincenzo1} has this fault--tolerant
property only for errors that occur {\em between} its successive quantum gate 
operations. Errors 
{\em during} gate operation that are needed to implement the error correcting
network will still produce many errors, as an error in 
a two--bit gate usually leads to two--bit errors. It is, fortunately, 
possible to improve this network such that these errors are also dealt 
with in a fault--tolerant way. This is an important improvement because 
we are more likely to experience errors during gate operation than 
in-between, as the interval between successive gates can be made very 
small compared to the gate operation time. 

Before we present our improved protocol for fault--tolerant error correction, 
we briefly review the procedure given in \cite{DiVincenzo1} and show that it
fails if errors occur during gate operation. The network under consideration (see fig. 1)
performs the error correction for a five--bit, single error correcting code 
\cite{Laflamme1,Bennett1}  with the code--words

\begin{eqnarray}
	|\tilde{0}\rangle & = & |C_0\rangle + |C_1\rangle \\
	|\tilde{1}\rangle & = & |C_0\rangle - |C_1\rangle
	\label{1}
\end{eqnarray}
where

\begin{eqnarray}
|C_0\rangle & = & |00000\rangle + |11000\rangle + |01100\rangle + |00110\rangle
		\nonumber \\
            & + & |00011\rangle + |10001\rangle - |10100\rangle - |01010\rangle
		\nonumber\\
            & - & |00101\rangle - |10010\rangle - |01001\rangle - |11110\rangle
		\nonumber\\ 
            & - & |01111\rangle - |10111\rangle - |11011\rangle - |11101\rangle
	\label{2}
\end{eqnarray}
and $|C_1\rangle$ being the state where each qubit is inverted with respect 
to $|C_0\rangle$. The fault--tolerant error correcting network for this code 
is presented in fig. 1 \cite{DiVincenzo1}. The 
incoming encoded state is represented by the top five lines. The lower 
four lines represent the error syndrome and are prepared in a known state. 
It should be noted that, to ensure fault--tolerant operation, each line (qubit) 
of the error syndrome actually consists of four separate qubits, 
initially prepared in a state with zero parity of the form

\begin{equation}
	|\Psi\rangle = \sum_{n,n\cdot \overline{1}=0} |n\rangle\;\; ,
	\label{4}
\end{equation}
where $\overline{1} = (1111)$ and $n\cdot \overline{1}$ is the bitwise 
product modulo $2$ \cite{Shor3}. That means that $|\psi\rangle$ is the 
equally weighted superposition of all four-qubit states of even parity. 
The action of the four CNOT operations on a qubit of the syndrome then
has to be rearranged as indicated in fig. 2. This ensures that, after an 
error in one of the CNOT's, 
\begin{figure}[hbt]
\newcounter{cms}
\setlength{\unitlength}{.6mm}
\begin{picture}(70,100)
\large
\put(5,90){\makebox(0,0)[c]{bit $0$}}
\put(5,80){\makebox(0,0)[c]{bit $1$}}
\put(5,70){\makebox(0,0)[c]{bit $2$}}
\put(5,60){\makebox(0,0)[c]{bit $3$}}
\put(5,50){\makebox(0,0)[c]{bit $4$}}
\put(10,30){\makebox(0,0)[c]{$a_0$}}
\put(10,20){\makebox(0,0)[c]{$a_1$}}
\put(10,10){\makebox(0,0)[c]{$a_2$}}
\put(10,0){\makebox(0,0)[c]{$a_3$}}
\normalsize
\thicklines
\multiput(15,-.1)(0,10){4}{\line(1,0){115}}
\multiput(15,.1)(0,10){4}{\line(1,0){115}}
\multiput(15,89.9)(0,-10){5}{\line(1,0){50}}
\multiput(15,90.1)(0,-10){5}{\line(1,0){50}}
\multiput(70,89.9)(0,-10){5}{\line(1,0){50}}
\multiput(70,90.1)(0,-10){5}{\line(1,0){50}}
\footnotesize
\multiput(67.5,90)(0,-10){5}{\circle{5}}
\multiput(67.5,90)(0,-10){5}{\makebox(0,0)[c]{{\bf R}}}
\multiput(122.5,90)(0,-10){5}{\circle{5}}
\multiput(122.5,90)(0,-10){5}{\makebox(0,0)[c]{{\bf R}}}
\normalsize
\multiput(125,89.9)(0,-10){5}{\line(1,0){5}}
\multiput(125,90.1)(0,-10){5}{\line(1,0){5}}
\multiput(20,90)(10,-10){5}{\circle*{2.5}}
\multiput(75,90)(10,-10){5}{\circle*{2.5}}
\put(19.9,90){\line(0,-1){92.5}}
\put(29.9,80){\line(0,-1){72.5}}
\put(39.9,70){\line(0,-1){72.5}}
\put(49.9,60){\line(0,-1){42.5}}
\put(59.9,50){\line(0,-1){42.5}}
\put(20.1,90){\line(0,-1){92.5}}
\put(30.1,80){\line(0,-1){72.5}}
\put(40.1,70){\line(0,-1){72.5}}
\put(50.1,60){\line(0,-1){42.5}}
\put(60.1,50){\line(0,-1){42.5}}
\put(74.9,90){\line(0,-1){62.5}}
\put(84.9,80){\line(0,-1){62.5}}
\put(94.9,70){\line(0,-1){62.5}}
\put(104.9,60){\line(0,-1){62.5}}
\put(114.9,50){\line(0,-1){52.5}}
\put(75.1,90){\line(0,-1){62.5}}
\put(85.1,80){\line(0,-1){62.5}}
\put(95.1,70){\line(0,-1){62.5}}
\put(105.1,60){\line(0,-1){62.5}}
\put(115.1,50){\line(0,-1){52.5}}
\put(20,0){\circle{5}}
\put(20,20){\circle{5}}
\put(30,10){\circle{5}}
\put(40,0){\circle{5}}
\put(40,30){\circle{5}}
\put(50,20){\circle{5}}
\put(60,10){\circle{5}}
\put(60,30){\circle{5}}
\put(20,0){\circle{5.2}}
\put(20,20){\circle{5.2}}
\put(30,10){\circle{5.2}}
\put(40,0){\circle{5.2}}
\put(40,30){\circle{5.2}}
\put(50,20){\circle{5.2}}
\put(60,10){\circle{5.2}}
\put(60,30){\circle{5.2}}
\multiput(75,30)(10,-10){4}{\circle{5}}
\multiput(85,30)(10,-10){4}{\circle{5}}
\multiput(75,30)(10,-10){4}{\circle{5.2}}
\multiput(85,30)(10,-10){4}{\circle{5.2}}
\put(130,-3){\line(0,1){36}}
\put(140,-3){\line(0,1){36}}
\put(130,-2.7){\line(1,0){10}}
\put(130,32.7){\line(1,0){10}}
\put(130.2,-3){\line(0,1){36}}
\put(140.2,-3){\line(0,1){36}}
\put(130,-2.9){\line(1,0){10}}
\put(130,32.9){\line(1,0){10}}
\footnotesize
\put(135,29){\makebox(0,0)[c]{S}}
\put(135,25){\makebox(0,0)[c]{y}}
\put(135,21){\makebox(0,0)[c]{n}}
\put(135,17){\makebox(0,0)[c]{d}}
\put(135,13){\makebox(0,0)[c]{r}}
\put(135,9){\makebox(0,0)[c]{o}}
\put(135,5){\makebox(0,0)[c]{m}}
\put(135,1){\makebox(0,0)[c]{e}}
\end{picture}\\[.5cm]
Fig.1: Fault--tolerant network given in \protect\cite{DiVincenzo1}: 
{\bf R} describes a one bit rotation which takes 
$|0\rangle \rightarrow (|0\rangle + |1\rangle)/\sqrt{2}$ and  
$|1\rangle \rightarrow (|0\rangle - |1\rangle)/\sqrt{2}$. 
An encircled cross denotes a NOT operation, while a small circle 
denotes a control bit. The first five qubits are encoded in the 
superposition of $|C_0\rangle$ and $|C_1\rangle$ given in 
eq. (\protect\ref{2}), while each of the last four `lines' represents 
four qubits initially in the state $|\Psi\rangle$ as in eq. (\ref{4}). 
At the end the error syndrome is obtained by performing a measurement 
on these $16$ qubits after which appropriate correction is applied to 
the first five qubits.

no back-action of 
errors takes place which would otherwise lead to multiple errors in the 
outgoing state \cite{Shor3}. One might think that this task could also 
be achieved 
with the initial syndrome state $|\Psi\rangle = |0000\rangle$ instead 
of the one given in eq. (\ref{4}). This is, however, not so, as then 
different code--words would lead to different states of $|\Psi\rangle$,
which would then enable us to single out one superposition state of the 
code from a measurement of the state resulting from $|\Psi\rangle$. 
The state eq. (\ref{4}), in contrast, contains only information about 
its parity. It can now be checked that the network presented in fig. 1 is
fault--tolerant if errors occur {\em between} operations of its quantum 
gates \cite{DiVincenzo1}. In table 1 we present the possible syndromes 
and the related errors, $X_i$ (amplitude error on the $i$--th bit), $Y_i$ 
(phase error on the $i$--th bit), and $Z_i = X_i\,Y_i$.

We now show by means of an example that one error during the operation 
of a CNOT can lead to two errors 
\end{figure}
\begin{figure}[hbt]
\setlength{\unitlength}{.6mm}
\begin{picture}(100,40)
\thicklines
\put(10,15){\line(1,0){40}}
\put(10,15.2){\line(1,0){40}}
\multiput(15,15)(10,0){4}{\circle{5}}
\multiput(15,15)(10,0){4}{\circle{5.2}}
\multiput(14.9,12.5)(10,0){4}{\line(0,1){25}}
\multiput(15.1,12.5)(10,0){4}{\line(0,1){25}}
\Huge
\put(65,15){\makebox(0,0)[c]{\boldmath $\Rightarrow$ \unboldmath}}
\multiput(77,0)(0,10){4}{\line(1,0){40}}
\multiput(77,0.2)(0,10){4}{\line(1,0){40}}
\multiput(82,30)(10,-10){4}{\circle{5}}
\multiput(82,30)(10,-10){4}{\circle{5.2}}
\put(81.9,27.5){\line(0,1){10}}
\put(91.9,17.5){\line(0,1){20}}
\put(101.9,7.5){\line(0,1){30}}
\put(111.9,-2.5){\line(0,1){40}}
\put(82.1,27.5){\line(0,1){10}}
\put(92.1,17.5){\line(0,1){20}}
\put(102.1,7.5){\line(0,1){30}}
\put(112.1,-2.5){\line(0,1){40}}
\end{picture}\\[.5cm]
Fig. 2: This diagram represents how the controlled NOTs from fig.1. are 
to be applied in order to avoid back--action of errors.
\end{figure}
\begin{center}
\renewcommand{\arraystretch}{1.}
\begin{tabular}{|c|c|}\hline
\,\,\,\,Error\,\,\,\, & \,\,\,Syndrome\,\,\, \\ 
                  & $a_0,a_1,a_2,a_3$\\ \hline\hline
$X_0$ & $0101$ \\ \hline
$X_1$ & $0010$ \\ \hline
$X_2$ & $1001$ \\ \hline
$X_3$ & $0100$ \\ \hline
$X_4$ & $1010$ \\ \hline
$Y_0$ & $1000$ \\ \hline
$Y_1$ & $1100$ \\ \hline
$Y_2$ & $0110$ \\ \hline
$Y_3$ & $0011$ \\ \hline
$Y_4$ & $0001$ \\ \hline
$Z_0$ & $1101$ \\ \hline
$Z_1$ & $1110$ \\ \hline
$Z_2$ & $1111$ \\ \hline
$Z_3$ & $0111$ \\ \hline
$Z_4$ & $1011$ \\ \hline
\end{tabular}
\end{center}
Table 1: All possible single bit errors and their corresponding 
syndromes that may occur in the network of fig. 1 are listed. $X_i$
denotes an amplitude error in bit $i$, $Y_i$ a phase error in bit $i$
and $Z_i=X_i Y_i$. All syndromes are different so that it is possible 
to correct a general single bit error.\\
in the `corrected' state in the scheme
presented in fig. 1. Assume that an error during the CNOT--operation 
between bit $0$ and $a_3$ has an effect as if there was an amplitude 
error in both bit $0$ and bit $a_3$ (in general the effect will be a
superposition of many possible two-bit errors). Then according to table 1 
the error syndrome would indicate an 
amplitude error $X_3$ which would subsequently be `corrected' and the 
outgoing state then has two amplitude errors in bits $0$ and $3$!
A state with two errors, however, cannot be dealt with by subsequent 
error correction steps which would, in actual fact, add even more 
errors to the state. Therefore the error correction procedure in 
fig. 1 cannot be regarded as fault-tolerant if errors occur {\em during} 
the gate operation. This is an important shortcoming because most errors 
will occur during the quantum gate operation and not in between,
as the time delay between successive quantum gates can be made 
much smaller than the time required to perform a quantum gate.
   
In the following we will show that the error correction 
scheme of DiVincenzo and Shor \cite{DiVincenzo1} which we have discussed 
above can be improved in order to make sure that also errors {\em during} 
quantum gate operation can be dealt with fault-tolerantly. To achieve 
this we have to repeat the generation of the above error syndrome 
{\em conditional} on the result of the first syndrome before we decide on 
the error correction step itself. Due to the additional information 
introduced by this conditional repetition we are now able 
to treat errors that occurred during gate operations. This is possible 
because although an error that occurs during a quantum gate introduces a 
two-bit error, one of the errors is in the syndrome. The error in the 
syndrome, however, does not propagate as each
\begin{center}
\begin{tabular}{|c|c|}\hline
\,\,\,\,Syndrome\,\,\,\, & Action \\ \hline\hline
$S_1=0$             & No correction \\ \hline
$S_1\neq 0$         & \,\,\,Generate another syndrome and\,\,\, \\ 
                    & correct error indicated by it \\ \hline
\end{tabular}
\end{center}
Table 2: The possible results for the syndrome $S_1$ after at
most one error at an arbitrary position in the network in fig. 3 
are shown together with the appropriate action that has to be taken for 
each of the outcomes.\\

syndrome is measured before 
the next one is produced. In table 2 the two possible outcomes for the 
first syndrome are shown together with the appropriate action that has 
to be taken. Using table 2 it is now simple to check that the network
in fig. 1 together with the conditional generation of the error syndrome
is capable of fault-tolerant error correction even if an error occurs 
at an {\em arbitrary} position in the error correcting network. 
It should be noted that we only produce another syndrome if there was
an error either in the incoming state or during the error correction network.
This, however, means that an error in the construction of the additional
syndrome would be a second order effect which we neglect here as we
only deal with single error correcting codes. It should also be noted
that it is important that the second syndrome is produced conditional on
the first one. If we would generate two syndromes from the outset then
we could obtain ambiguous results exactly in the case where the first 
syndrome does not indicate an error but the second one does. This would
then require the generation of yet a third syndrome conditional on the
outcome of the first two syndromes. 

We can summarize the result of our error correction protocol
with conditional generation of error syndromes by:\\
{\em If the incoming state is error--free 
and only one error occurs at an arbitrary position during the network 
operation, the outgoing state has at most one error. If the incoming 
state has one error and no further error occurs during the error 
correction then the outgoing state will be corrected perfectly.} 

This allows fault-tolerant error correction if at most one error occurs
in the incoming state and the error correction step together. This is 
achieved using the idea of conditional construction of error syndromes 
in the 'fault--tolerant' error correction network given by 
\cite{DiVincenzo1}. The scheme of \cite{DiVincenzo1} alone, as we saw, 
could not cope with the errors occurring during the gate operation. As 
these errors can be expected to be predominant this is an important 
shortcoming of the procedure. The error correction protocol with conditional
generation of the error syndrome, as presented here, however, is 
fault--tolerant even if an error occurs during the gate operation, and 
hence can fault--tolerantly correct for a general single error at 
{\em arbitrary} position. The result that conditional error syndromes
can be used to treat errors during gate operation is also interesting 
because it shows that it is not always necessary to implement the 
quantum gates in a fault-tolerant way. This can simplify
the construction of the error correction networks because although
for special, practically important, errors there exist efficient 
fault-tolerant implementations of quantum gates \cite{Cirac2} in 
general the fault-tolerant implementation of quantum gates can be 
complicated.  \\[1.cm]
{\bf Acknowledgements}\\[.25cm]
One of us (MBP) would like to thank D.P. DiVincenzo for useful discussions.
This work was supported by the European Community, the UK Engineering and 
Physical Sciences Research Council and by a Feodor-Lynen grant of the 
Alexander von Humboldt foundation.

\end{document}